\documentclass[
preprint,
superscriptaddress,
nobibnotes,
amsmath,amssymb,
aps,
prm,
]{revtex4-2}

\usepackage{graphicx}
\usepackage{dcolumn}
\usepackage{bm}
\usepackage{subfig}
\usepackage{booktabs}
\usepackage{multirow}
\usepackage{colortbl}
\usepackage{xcolor}
\usepackage{ulem}
\begin{document}

\title{Monolayer control of spin-charge conversion in van der Waals heterostructures}

\author{Khasan Abdukayumov}
\affiliation{Université Grenoble Alpes, CEA, CNRS, IRIG-SPINTEC, 38000 Grenoble, France}

\author{Oliver Paull}
\affiliation{Laboratoire Albert Fert, CNRS, Thales, Université Paris-Saclay, 1 avenue Augustin Fresnel, 91767, Palaiseau France}

\author{Martin Mi\v{c}ica}
\affiliation{Laboratoire de Physique de l’Ecole normale sup\'{e}rieure, ENS, Universit\'{e} PSL, CNRS, Sorbonne Universit\'{e}, Universit\'{e} de Paris, 24 rue Lhomond, 75005 Paris, France}

\author{Fatima Ibrahim}
\author{Libor Voj\'{a}\v{c}ek}
\affiliation{Université Grenoble Alpes, CEA, CNRS, IRIG-SPINTEC, 38000 Grenoble, France}

\author{Adrien Wright}
\affiliation{Laboratoire de Physique de l’Ecole normale sup\'{e}rieure, ENS, Universit\'{e} PSL, CNRS, Sorbonne Universit\'{e}, Universit\'{e} de Paris, 24 rue Lhomond, 75005 Paris, France}

\author{Sylvain Massabeau}
\affiliation{Laboratoire Albert Fert, CNRS, Thales, Université Paris-Saclay, 1 avenue Augustin Fresnel, 91767, Palaiseau France}

\author{Federico Mazzola} 
\affiliation{Department of Molecular Sciences and Nanosystems, Ca’Foscari University of Venice, Venice, Italy}

\author{Vincent Polewczyk}
\author{Cyriack Jego}
\author{Rahul Sharma}
\author{C\'eline Vergnaud}
\author{Alain Marty}
\author{Isabelle Gomes de Moraes}
\affiliation{Université Grenoble Alpes, CEA, CNRS, IRIG-SPINTEC, 38000 Grenoble, France}

\author{Abdelkarim Ouerghi}
\affiliation{Université Paris-Saclay, CNRS, Centre de Nanosciences et de Nanotechnologies, 91120,
Palaiseau, France}

\author{Hanako Okuno}
\affiliation{Université Grenoble Alpes, CEA, IRIG-MEM, 38000 Grenoble, France}

\author{Anupam Jana}
\author{Indrani Kar}
\author{Jun Fuji}
\author{Ivana Vobornik}
\affiliation{CNR-IOM Laboratorio TASC, I-34149 Trieste, Italy}

\author{Jing Li}
\affiliation{Université Grenoble Alpes, CEA-LETI, 38000 Grenoble, France}

\author{Fr\'ed\'eric Bonell}
\affiliation{Université Grenoble Alpes, CEA, CNRS, IRIG-SPINTEC, 38000 Grenoble, France}

\author{Mairbek Chshiev}
\affiliation{Université Grenoble Alpes, CEA, CNRS, IRIG-SPINTEC, 38000 Grenoble, France}
\affiliation{Institut Universitaire de France, Paris, 75231 France}

\author{Manuel Bibes}
\author{Jean-Marie George}
\author{Henri Jaffrès}
\affiliation{Laboratoire Albert Fert, CNRS, Thales, Université Paris-Saclay, 1 avenue Augustin Fresnel, 91767, Palaiseau France}

\author{Sukhdeep Dhillon}
\affiliation{Laboratoire de Physique de l’Ecole normale sup\'{e}rieure, ENS, Universit\'{e} PSL, CNRS, Sorbonne Universit\'{e}, Universit\'{e} de Paris, 24 rue Lhomond, 75005 Paris, France}
\email{Sukhdeep.Dhillon@ens.fr}

\author{Matthieu Jamet}
\affiliation{Université Grenoble Alpes, CEA, CNRS, IRIG-SPINTEC, 38000 Grenoble, France}
\email{matthieu.jamet@cea.fr}



\date{\today}

\begin{abstract}

The diversity of 2D materials and their van der Waals (vdW) stacking presents a fertile ground for engineering novel multifunctional materials and quantum states of matter. This permits unique opportunities to tailor the electronic properties of vdW heterostructures by the insertion of only a single 2D material layer. However, such vdW materials engineering at the atomic scale has yet to be investigated for spin-charge interconversion phenomena. Here, we report on the control of these effects at the monolayer level, where drastic increase in intensity and change in sign of THz spintronic emission are demonstrated by inserting a single layer of MoSe$_2$ between PtSe$_2$ and graphene in a fully epitaxial, large area stacked structure. By using a combination of spin and angle resolved photoemission and density functional theory to reveal the electronic and spin structures, we illustrate two different mechanisms relying on charge transfer and electronic hybridization for the formation of Rashba states, which are responsible for spin-charge conversion and hence the THz spintronic emission. These findings open new pathways to design, at the atomic scale, efficient THz spintronic emitters made of 2D materials and other spintronic devices based on spin-charge interconversion phenomena. 

\end{abstract}

\maketitle

\textit{Introduction - }The field of two-dimensional (2D) van der Waals (vdW) heterostructures has revealed a whole new world of exotic materials hosting a wide variety of physical properties \cite{Geim2013,Andrei2021}. In 2D transition metal dichalcogenides (TMDs), spin-charge interconversion has emerged as a promising avenue for spintronics information processing and storage by virtue of their atomically thin nature and layer-dependent electronic and magnetic properties \cite{Galceran2021,Sierra2021,Abdukayumov}. This is a core phenomenon in the field of spintronics, wherein a spin current can be converted into a charge current and vice versa through a number of mechanisms that are based on spin-orbit coupling \cite{RojasSanchez2019}. This interconversion can be the result of the Rashba-Edelstein effect \cite{RojasSanchez2013}, the spin-Hall effect \cite{Valenzuela2006} or more recently the orbital-Hall effect \cite{Go2018} (both direct and inverse). The spin-Hall effect, being a bulk effect, occurs in materials where there is strong spin-orbit coupling, typically requiring particular materials with heavy elements. On the other hand, the Rashba-Edelstein effect is a surface/interface effect and can be induced artificially by introducing an interface in a system to break inversion symmetry. It opens up spin-charge interconversion to a myriad of material combinations and, more particularly, to vdW heterostructures where interfaces are atomically sharp and well defined \cite{Yang2022,Wang2022}. Initially, exfoliation methods from bulk samples produced the highest quality 2D monolayer samples at the expense of being limited to small areas on the order of micrometres. Similarly, for bilayer 2D heterostructures, advanced exfoliation and stacking methods have been developed \textit{in} and \textit{ex vacuo}. Only more recently have growth methods such as molecular beam epitaxy (MBE) and chemical vapor deposition (CVD) have produced high crystalline quality 2D materials and vdW heterostructures through the optimization of surface preparation and specific growth recipes \cite{Cai2018,Dau2018,Dimoulas2015,King2020,Vergnaud2020,Dau2019}. Throughout this development in synthesis and heterostructure design, it has become apparent that the interface between van der Waals materials permits considerably more versatility compared to traditional epitaxy owing to the alleviated requirement of lattice matching of epilayers. With the mitigation of this requisite, additional methods to add and tune properties, such as twisted-angle bilayers \cite{Carr2017}, have become feasible. The ability to epitaxially grow 2D TMDs in a high-quality heterostructure on large area has been demonstrated by a limited number of research groups \cite{gong_vertical_2014,Guillet_Epitaxial_2023, Khalil2022,Guillet2024,Aretouli2016,Kesilebiek2020,Ohtake2021}. \\
Here, we demonstrate the spin-charge conversion (SCC) at PtSe$_2$/Gr and PtSe$_2$/MoSe$_2$ interfaces grown by molecular beam epitaxy on large area, and show that the sign and amplitude of the SCC can be drastically modified by changing the adjacent material from graphene to MoSe$_2$ hence allowing monolayer control of SCC. It has been previously shown that the spin texture in PtSe$_2$ displays a helical character which evolves with the number of layers \cite{yan_high_2017,yao_direct_2017}. To design a spintronic device based on the Rashba-induced helical spin texture of PtSe$_2$, accurate control over this spin texture must therefore be established. We employ spin and angle-resolved photoelectron emission spectroscopy (Spin-ARPES) and THz emission spectroscopy to probe the spin-charge conversion characteristics and reveal that inserting only a single layer of MoSe$_2$ between PtSe$_2$ and graphene induces a sign change and stronger SCC. First-principles calculations indeed demonstrate two different mechanisms at the origin of the spin texture: charge transfer at the PtSe$_2$/Gr interface and electronic hybridization at the PtSe$_2$/MoSe$_2$ van der Waals interface. This work shows the strength of vdW materials to fine-tune spin texture and SCC at the atomic scale in future spintronic devices thanks to well-defined interfaces.\\ 

\begin{figure}[ht!!]
    \centering
    \includegraphics[width=\linewidth]{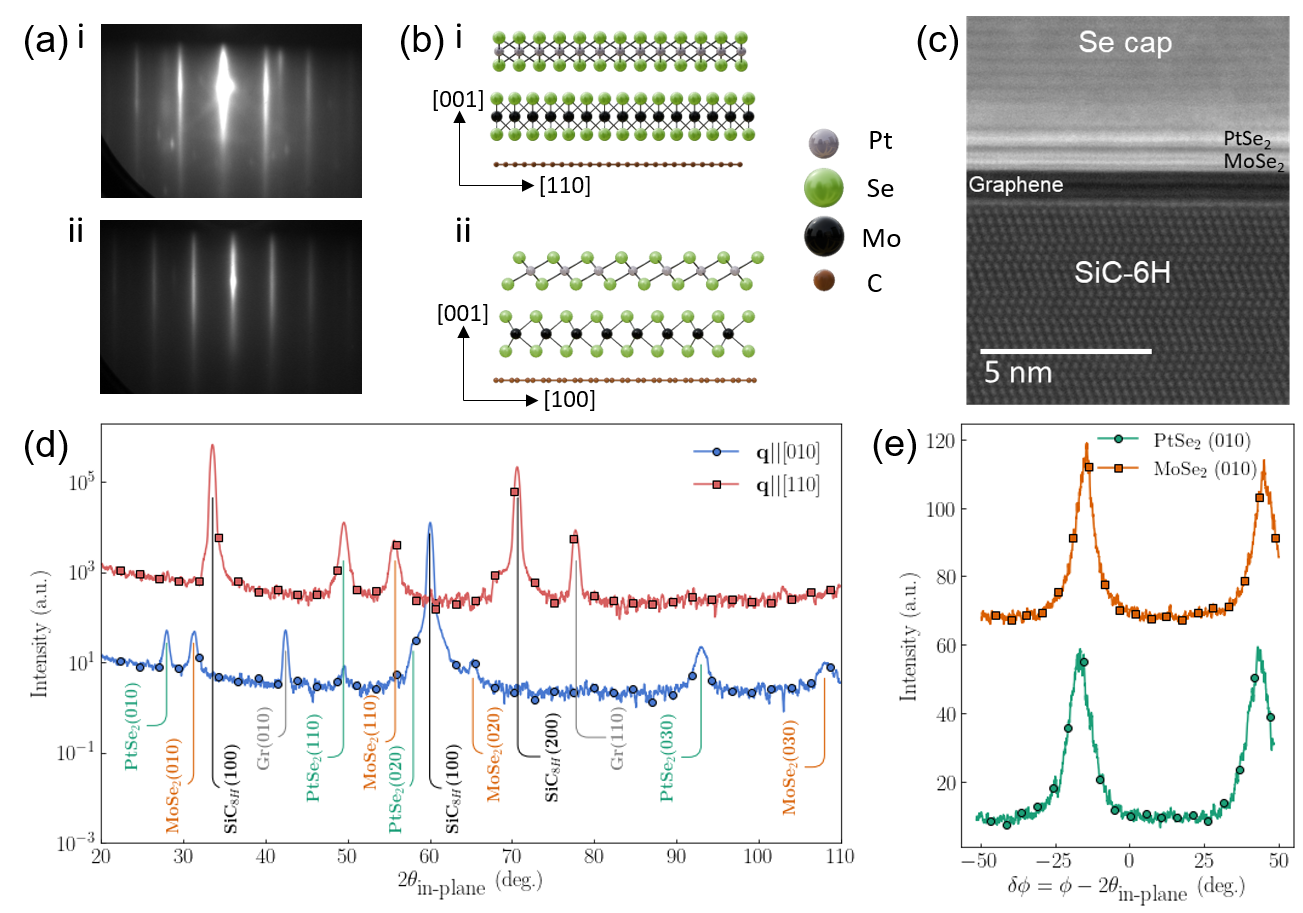}
    \caption{Growth and characterization of the PtSe$_2$/MoSe$_2$/Gr bilayer. (a) RHEED images of (i) MoSe$_2$ monolayer and (ii) PtSe$_2$ monolayer along [100]. (b) Atomic model of the layers viewed from the (i) [100] and (ii) [110] directions. (c) Transmission electron microscopy image in cross section of the vdW heterostructure capped with amorphous Se. (d) In-plane radial x-ray diffraction measurements with the momentum transfer $\mathbf{q}$ parallel to the [010] (blue line and circles) and [110] (red line and squares) directions. (e) In-plane azimuthal x-ray diffraction of the PtSe$_2$ (010) peak (green line and circles) and MoSe$_2$ (010) peak (orange line and squares).}
    \label{fig:growth}
\end{figure}

\textit{Material growth and characterization - }Bilayers of MoSe$_2$ and PtSe$_2$ were grown on a graphene/6H-SiC(0001) substrate using a custom MBE system \cite{Suppl_Mater,Pallecchi2014,Abdukayumov}. Reflection high-energy electron diffraction (RHEED) was used to monitor the layer-by-layer growth \textit{in-situ}, displaying diffraction rods for MoSe$_2$ (Fig.\ref{fig:growth}(a)i) and PtSe$_2$ (Fig.\ref{fig:growth}(a)ii). We observe very thin and continuous diffraction lines showing the high crystalline quality of the vdW materials and the epitaxial relationship between MoSe$_2$ and PtSe$_2$. 
The layer stacking is shown in Fig.\ref{fig:growth}(b)i (viewed along [100]) and Fig.\ref{fig:growth}(b)ii (viewed along [110]) and depicts epitaxial 2D monolayers of PtSe$_2$ and MoSe$_2$ on graphene. The corresponding cross section transmission electron microscopy \cite{Suppl_Mater} image is shown in Fig.\ref{fig:growth}(c) where we clearly distinguish the TMD layers and graphene.\\
To verify the epitaxy and single-crystalline nature of the 2D heterostructure, in-plane X-ray diffraction measurements (XRD) were performed \cite{Suppl_Mater}. In Fig.\ref{fig:growth}(d), radial scans confirm that the [010] and [110] crystal directions for MoSe$_2$ and PtSe$_2$ are epitaxially aligned with those of graphene (\textit{i.e.} with a twist angle of 0$^{\circ}$). The azimuthal XRD scans in Fig.\ref{fig:growth}(e) exhibit peaks separated by 60 degrees as expected for hexagonal symmetry, and demonstrate the high-quality single-crystal nature of the 2D layers. \\

\begin{figure}[ht!!]
    \centering
    \includegraphics[width=\linewidth]{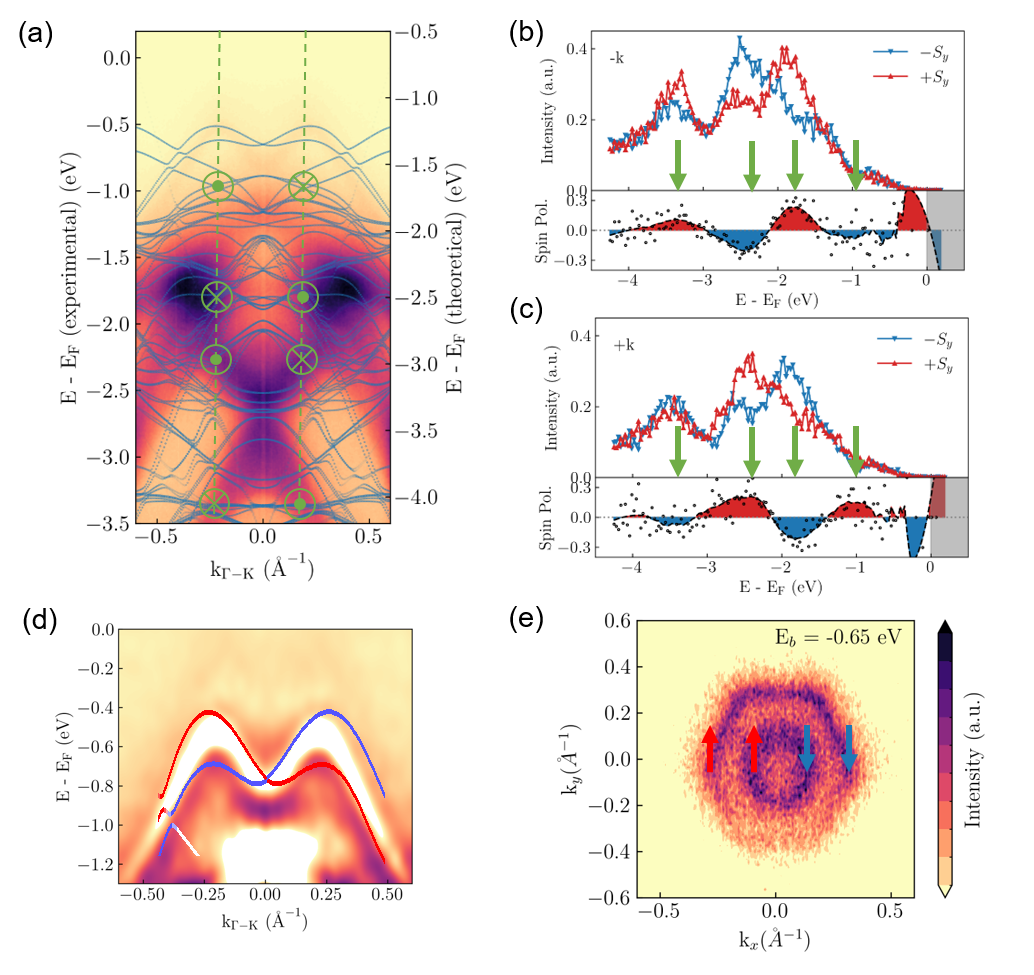}
    \caption{spin-ARPES of the PtSe$_2$/MoSe$_2$/Gr heterostructure. (a) Energy sweep along the $\Gamma-$K direction (left axis) with overlaid band structure calculated by DFT (blue, right axis). Spin-resolved energy distribution curves (s-EDCs) at (b) -0.23 \AA$^{-1}$ and (c) +0.23 \AA$^{-1}$ corresponding to the vertical dashed lines in (a). Spin textures are reported in (a) according to the spin polarization along $y$ in (b) and (c) at specific energies indicated by vertical green arrows. (d) Second derivative of the raw Spin-ARPES data highlighting the spin-splitting of the sombrero band. The band positions correspond to the dark areas (negative value) on both sides of the bright one (positive value). The red (resp. blue) line corresponds to the calculated band with spin pointing along $+\hat{y}$ (resp. $-\hat{y}$). (e) Constant-energy map in the hybridized sombrero band along with the calculated spin texture at E$_b$=-0.65 eV.}
    \label{fig:arpes}
\end{figure}

To elucidate the spin texture of this bilayer, Spin-ARPES was used. The photoemission experiments were performed at the low-energy end-station of the Advanced Photoemission Experiment (APE) beamline, at the synchrotron radiation source Elettra \cite{Bigi2017} (Trieste). The energy and momentum resolutions were better than 12 meV and 0.02 \AA$^{-1}$, respectively. The spin-integrated energy versus momentum spectra along the $\Gamma$-K high symmetry direction of the Brillouin zone are shown in Fig.~\ref{fig:arpes}(a), along with the theoretical electronic structure calculated by density functional theory-based approach (DFT, blue lines) \cite{Suppl_Mater}. An overall good agreement between the experimental measurements and calculations was observed and the main spectroscopic features are captured by the DFT. This includes, in agreement with previous calculations \cite{xiang_tunable_2019}, a sombrero-like dispersion, with a maximum located $\approx$0.5 eV below the Fermi level. Such a sombrero-dispersion, according to our DFT calculations, is spin-split, featuring spin-momentum locking with Rashba-like physics. Therefore, the spins change sign from positive to negative values of momenta ($\pm k$). The spin-momentum locking is directly demonstrated by the Spin-ARPES measurements (see Fig.\ref{fig:arpes}(b-c) for selected energy-distribution curves resolved in ($\sigma_y$) which show strong signals with $+k$ and $-k$ connected by the enforcement of time-reversal symmetry in the system. We notice that the spin-splitting of the sombrero-dispersion in Fig.\ref{fig:arpes}(b) and (c) is not immediately resolved by Spin-ARPES, probably owing to a combination of photoemission matrix elements and experimental resolutions. However, the band edges can be identified by using the second derivative of the spectra in Fig.\ref{fig:arpes}(d), allowing us to set 0.15 eV as the upper value for the expected spin-splitting. This is highlighted by plotting the predicted spin-polarized bands (blue and red lines) calculated by DFT on the same figure. To better identify the fermiology of the sombrero-dispersion, we extracted a constant energy map (Fig.\ref{fig:arpes}(e)) at E$_{b}$=-0.65 eV with the calculated spin texture. The shape reflects the hexagonal symmetry of the crystal. Remarkably, compared to individual PtSe$_2$ and MoSe$_2$ monolayers, the top of the sombrero-dispersion is closer to the Fermi level and originates from interface hybridization between PtSe$_2$ and MoSe$_2$ as discussed in the following. We theoretically found the same sombrero-dispersion for the PtSe$_2$/WSe$_2$ bilayer \cite{Suppl_Mater} showing the general character of this hybridization. Further by selecting properly the 2D material couple, it could be possible to create a two-dimensional electron or hole gas at the interface between two semiconducting materials (i.e. hybridized band crossing the Fermi level), which is reminiscent of 2D electron gases in LaAlO$_3$/SrTiO$_3$ interfaces. It points at the important role of orbital hybridization to create artificially metallic interfacial systems \cite{Hwang2004}. \\

\begin{figure}[ht!!]
    \centering
    \includegraphics[width=\linewidth]{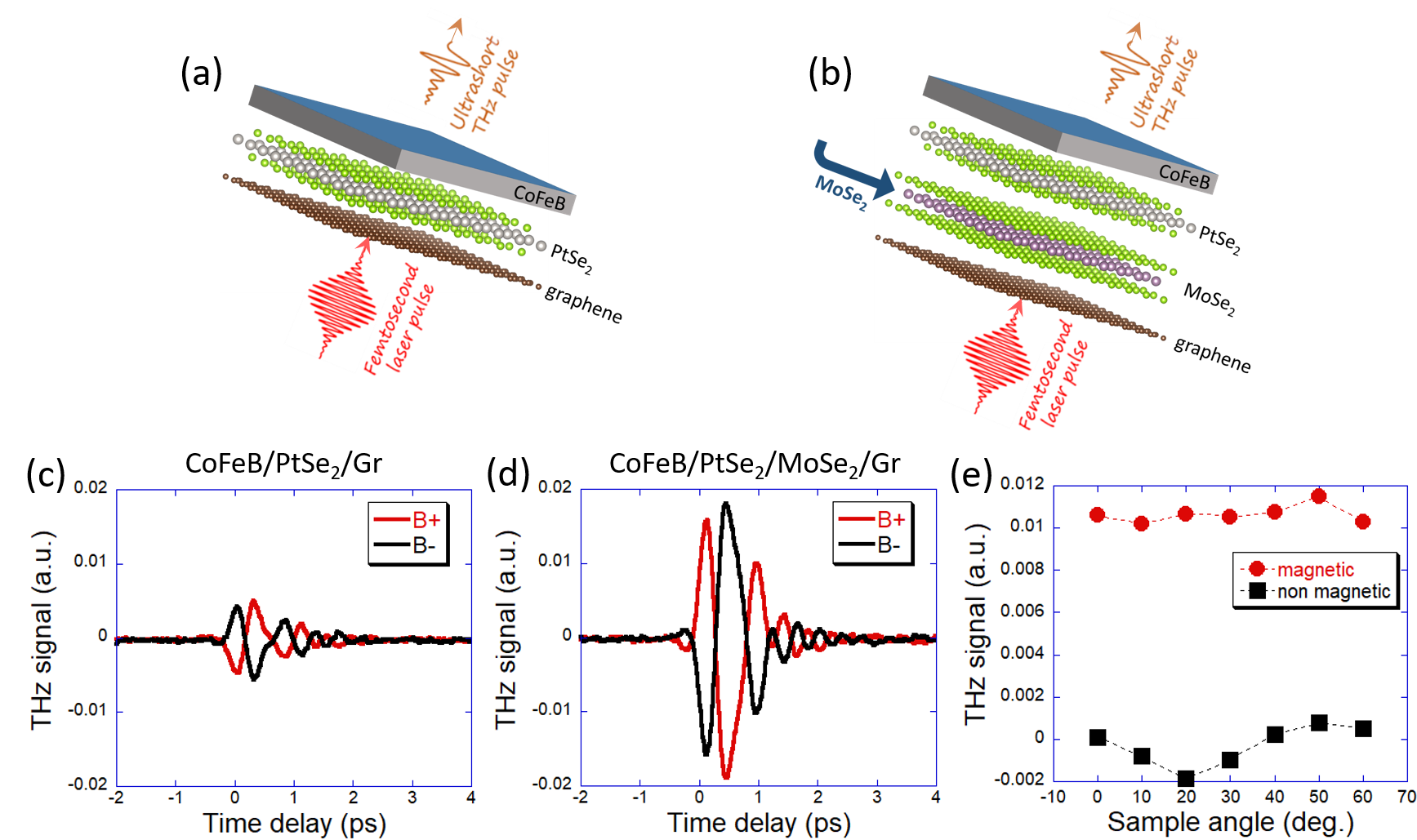}
    \caption{THz emission results. (a) and (b), schematic of femtosecond optical pumping and THz pulse emission for 1 ML PtSe$_2$/Gr and bilayer PtSe$_2$/MoSe$_2$/Gr systems, respectively. (c) and (d) THz electric field as a function  of time delay and magnetic field direction for PtSe$_2$/Gr and PtSe$_2$/MoSe$_2$/Gr systems, respectively. The external magnetic field (strong enough to saturate the CoFeB magnetization) is applied along an azimuthal angle of 0° (red) and 180° (black). (e) Angular dependence of the magnetic $S_M$ and non-magnetic $S_{NM}$ contributions for a fixed magnetic field direction  to the emitted THz electric field for PtSe$_2$/MoSe$_2$/Gr.}
    \label{fig:THz}
\end{figure}

\textit{THz measurements - }THz time domain spectroscopy experiments were carried out on the PtSe$_2$/MoSe$_2$ bilayer, and compared with those on 1 ML PtSe$_2$ epitaxially grown on graphene/SiC \cite{Suppl_Mater}. To perform these measurements, 3 nm of CoFeB and 4 nm of Al were deposited which naturally oxidizes into AlO$_x$. The sharp and high quality of the ferromagnet/TMD interface has been confirmed by Raman spectroscopy, x-ray photoemission spectroscopy (XPS) and scanning transmission electron microscopy (STEM)\cite{Abdukayumov}.
As schematically shown in Fig.~\ref{fig:THz}(a-b), the measurements were carried out with a femtosecond excitation (centre wavelength of 800 nm (1.55~eV)) for both samples in transmission mode, generating a spin current in the CoFeB layer that is converted into an ultrafast charge current in the 2D material, emitting a THz pulse by SCC. The sample was pumped from the substrate side as it showed stronger measured THz electric field. As presented in Fig.~\ref{fig:THz}(c) and (d), the bilayer PtSe$_2$/MoSe$_2$/Gr sample exhibits around 3.5 times stronger THz electric field than 1 ML of PtSe$_2$/Gr with opposite THz electric field polarity. To extract the magnetic contribution $S_M$ and non-magnetic contribution $S_{NM}$ from the THz signal $S_{THz}$, we use: $S_M=(S_{THz}(+B)-S_{THz}(-B))/2$ and $S_{NM}=(S_{THz}(+B)+S_{THz}(-B))/2$ \cite{Rongione2023}. The azimuthal angular dependence of the magnetic contribution $S_M$ in Fig.~\ref{fig:THz}(e) corresponds to the SCC and shows an isotropic behavior as expected for THz spintronic emission \cite{Rongione2023}. In Fig.~\ref{fig:THz}(e), for the non-magnetic contribution, $S_{NM}$, we find very weak angular dependence and do not recover the expected 6-fold symmetry from non-linear THz emission of 1 ML MoSe$_2$ owing to inversion symmetry breaking \cite{Khan2022}. This observation emphasizes the electronic hybridization of MoSe$_2$ with PtSe$_2$, which modifies the electronic band structure and symmetries of the system.


Moreover, due to its proximity to the Fermi level and low effective mass, the sombrero-like band is likely to contribute to SCC for the THz emission thus justifying the different intensity and sign of THz emission when compared to PtSe$_2$/Gr as discussed further below.\\

\begin{figure}[ht!!]
    \centering
    \includegraphics[width=\linewidth]{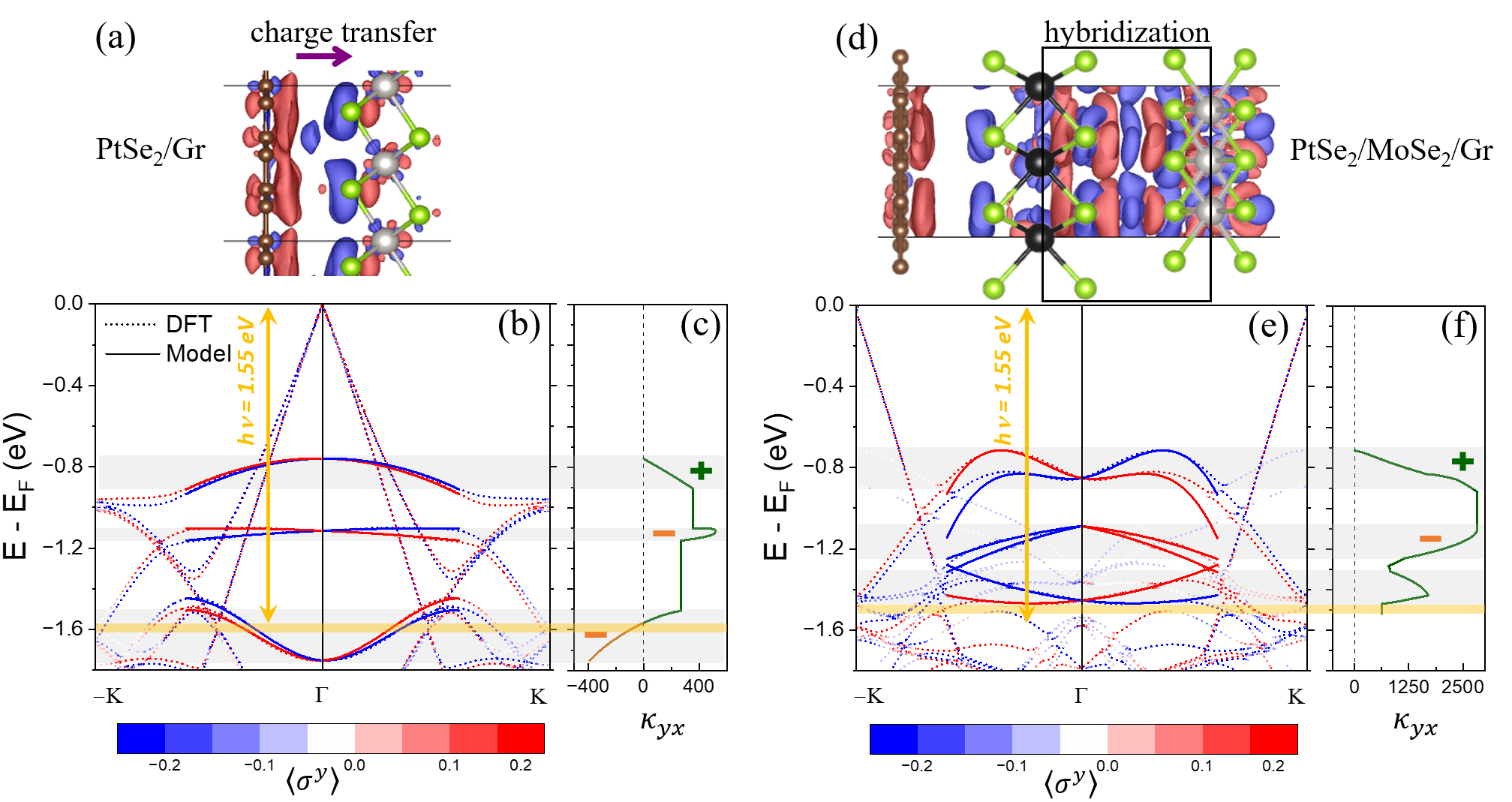}
    \caption{Crystal structure and calculated charge density in PtSe$_2$/Gr (a) and PtSe$_2$/MoSe$_2$/Gr (d) heterostructures. Blue (red) clouds represent accumulation (depletion) of electrons respectively using an isosurface of 10$^{-4}$ $e$/\AA$^3$ . The interaction mechanism in PtSe$_2$/Gr is mainly charge transfer whereas the overlap of charge clouds points toward a strong hybridization between PtSe$_2$ and MoSe$_2$. The band structure with spin projections along $y$-direction $\langle\sigma^{y}\rangle$ are shown for PtSe$_2$/Gr (b) and  PtSe$_2$/MoSe$_2$/Gr (e). The dotted lines are the DFT calculated bands whereas the solid lines are the fitted ones using the free-electron model described in the text. The SCC efficiency is quantified by the calculated $\kappa_{yx}$ tensor in arbitrary units where its cumulative profile is displayed for PtSe$_2$/Gr (c) and PtSe$_2$/MoSe$_2$/Gr (f). The negative sign corresponds to the SCC sign of Co/Pt.}
    \label{fig:theory}
\end{figure}

\textit{DFT calculations - }To understand in more details the role of MoSe$_2$ insertion between PtSe$_2$ and graphene on SCC, we have employed first-principles calculations to compare the PtSe$_2$/MoSe$_2$/Gr to the PtSe$_2$/Gr interface \cite{Suppl_Mater,Blochl1994,Kresse1993,Kresse1996a,Kresse1996b,Perdew1996,Bucko2010,Xiang2019}. The interaction across the interface is preliminarily described by the induced changes in the charge density shown in Fig.~\ref{fig:theory} (a) and (d) for PtSe$_2$/Gr and PtSe$_2$/MoSe$_2$/Gr heterostructures, respectively. From the planar averages \cite{Suppl_Mater}, the estimated charge transfer is 0.04$e$ from graphene to PtSe$_2$ while a less significant 0.007$e$ is transferred from MoSe$_2$ to PtSe$_2$. This implies that the dipoles at both interfaces have the same direction pointing towards the PtSe$_2$ layer with a smaller magnitude at the PtSe$_2$/MoSe$_2$ interface. Therefore, the change of SCC sign upon MoSe$_2$ insertion cannot be explained in a simple picture of reversing the interface dipole. However, it can be clearly seen in Fig.~\ref{fig:theory}(d) that the overlap of the charge clouds is more significant at this interface implying a strong hybridization between the PtSe$_2$ and MoSe$_2$ layers. In conclusion, the interaction mechanism in PtSe$_2$/MoSe$_2$/Gr governed by hybridization is different from PtSe$_2$/Gr that is driven by charge transfer. 
The hybridization strength is discerned from the calculated band structures in Fig.~\ref{fig:theory} (b) and (e) for PtSe$_2$/Gr and PtSe$_2$/MoSe$_2$/Gr, respectively. A weak hybridization is reflected through the preserved band structure features of both pristine graphene and PtSe$_2$, while the modified band features reveal the stronger hybridization between PtSe$_2$ and MoSe$_2$. Importantly, the inversion symmetry breaking imposed by the interface in both systems induces a Rashba splitting of the bands owing to the presence of spin-orbit interaction. This is depicted from the expected values of the spins projected on the $y$-direction $\langle\sigma^{y}\rangle$ along the -K $\rightarrow$ $\Gamma$ $\rightarrow$ K path in the Brillouin zone. To compare the SCC efficiency in PtSe$_2$/MoSe$_2$/Gr to PtSe$_2$/Gr, we proceed by fitting the calculated band structures to a free-electron model as described in the Supplemental Metarial \cite{Suppl_Mater,Vaz2019}. This permits to quantify the SCC efficiency characterized by the tensor $\kappa_{yx}$ integrated along the $k_x$ path \cite{Vaz2019}:

\begin{equation}
\kappa_{yx}\propto \int_{-k_{x}^{lim}}^{+k_{x}^{lim}} \langle \sigma^{y}
\rangle v_{g}^{x} 2 \pi |k| dk
\end{equation}

where $k_{x}^{lim}$ was chosen as 0.6$\times|\Gamma K|$ length, $v_{g}^{x}$=$\partial E / \partial k_x$ is the group velocity along $x$ and $|k|$ accounts for the growing Fermi line circumference with the distance from the $\Gamma$ point. Fig.~\ref{fig:theory} (c) and (f) show the cumulative $\kappa_{yx}$ values in arbitrary units arising from the highest Rashba-splitted valence bands in PtSe$_2$/Gr and PtSe$_2$/MoSe$_2$/Gr, respectively. We only focus here on the valence bands since the conduction bands show almost no spin polarization in both systems. From the energy profile of $\kappa_{yx}$, for the PtSe$_2$/Gr case the highest valence band has a weak positive contribution which is competing with the negative one of the lower two bands. On the other hand, the highest occupied band (sombrero hat) featuring the strong PtSe$_2$/MoSe$_2$ hybridization acquires a large Rashba splitting that drives the overall SCC to large positive values. This cannot be compensated by the negative contribution from the lower band. A detailed description of the bands' contribution to the SCC sign is provided in the Supplemental Material \cite{Suppl_Mater}. Since the exact position of the Fermi level is unknown and might be different for PtSe$_2$/Gr and PtSe$_2$/MoSe$_2$/Gr, we assume that it is positioned at zero energy in the band diagrams of Fig.~\ref{fig:theory}(b) and (e). The photon energy of the excitation laser ($\approx$1.55 eV) is indicated as an orange arrow in Fig.~\ref{fig:theory}(b) and (e) and sets the deepest valence bands involved in SCC. Considering the narrow energy windows close to the photon energy in orange in Fig.~\ref{fig:theory}(b) and (e), we theoretically obtain -150$\le \kappa_{yx}\le$0 for PtSe$_2$/Gr and $\kappa_{yx}\approx$600 for PtSe$_2$/MoSe$_2$/Gr which is in good qualitative agreement with experimental results in terms of SCC sign and relative intensity. 


\textit{Conclusion - }In conclusion, we demonstrate the growth of single crystalline van der Waals heterostructures over large areas using MBE and study SCC through THz spintronic emission. We show that inserting a single layer of MoSe$_2$ between graphene and PtSe$_2$ drastically changed the resulting SCC intensity and sign. DFT calculations supported by Spin-ARPES measurements pointed at two different mechanisms for the formation of the Rashba states responsible for SCC. In PtSe$_2$/Gr, the spin texture comes from charge transfer between PtSe$_2$ and graphene while it originates from the strong electronic hybridization between the two TMD layers in PtSe$_2$/MoSe$_2$/Gr as a consequence of the epitaxial stack. The calculation of SCC based on the band structure agrees well with the spintronic THz emission. This work demonstrates the great interest of 2D materials and their vdW heterostructures to fine-tune spin textures and SCC at the atomic scale thanks to well-defined vdW interfaces. They thus constitute promising systems to design compact and efficient spintronic THz emitters or other spintronic devices involving SCC.

\section*{Acknowledgements}

The authors acknowledge the support from the European Union’s Horizon 2020 research and innovation Programme under grant agreements No 881603 (Graphene Flagship), No 964735 (FET-OPEN EXTREME-IR) and  No 800945-NUMERICS-H2020-MSCA-COFUND-2017. O.P. and M.B. acknowledge financial support from ERC AdG "FRESCO" (\#833973). Spintec was
supported by the FLAG-ERA grant MNEMOSYN. The French National Research Agency (ANR) is acknowledged for its support through the ANR-18-CE24-0007 MAGICVALLEY, ANR-20-CE24-0015 ELMAX and ESR/EQUIPEX+ ANR-21-ESRE-0025 2D-MAG projects. The LANEF framework (No. ANR-10-LABX-0051) is acknowledged for its support with mutualized infrastructure. Part of the calculations were performed on computational resources provided by GENCI–IDRIS (Grant 2024-A0150912036).




\pagebreak

\end{document}